\documentclass[prl,showpacs,twocolumn, preprintnumbers,amsmath,amssymb]{revtex4}
\usepackage{amsmath}
\usepackage{amssymb}
\usepackage{graphicx}
\usepackage{dcolumn}

\newcommand{\bm}{\ensuremath{\mathbf{{m}}}}

\newcommand{\bj}{\ensuremath{\mathbf{{j}}}}

\newcommand{\Kn}{\ensuremath{{\rm Kn}}}

\newcommand{\RE}{\ensuremath{{\rm Re}}}

\newcommand{\La}{\ensuremath{{\rm La}}}
\newcommand{\Ma}{\ensuremath{{\rm Ma}}}
\newcommand{\AKS}{\ensuremath{{L_{\rm a}}}}
\newcommand{\Cv}{\ensuremath{{C}_{\rm V}}}
\newcommand{\kT}{\ensuremath{ \frac{\partial P}{\partial T } \Biggr|_{\rho}} }
\newcommand{\kTmean}{\ensuremath{ \frac{\partial P}{\partial T } \Biggr|_{\rhom}} }
\newcommand{\rhom}{\ensuremath{\overline{{\rho}}}}

\newcommand{\Tm}{\ensuremath{\overline{{T}}}}

\newcommand{\bx}{\ensuremath{\mathbf{x}}}

\newcommand{\bu}{\ensuremath{\mathbf{u}}}
\newcommand{\bv}{\ensuremath{\mathbf{v}}}

\newcommand{\blambda}{\mbox{\boldmath$\lambda$}}

\begin{document}
\title
{Thermodynamic theory of incompressible
hydrodynamics\footnote{Physical Review Letters, accepted for
publication (2005).}}
\author{Santosh Ansumali}
\affiliation {ETH-Z\"urich, Institute of Energy Technology,
CH-8092 Z\"urich, Switzerland} \email{ansumali@lav.mavt.ethz.ch}
\author{ Iliya V. Karlin}
\affiliation {ETH-Z\"urich,  Institute of Energy Technology,
CH-8092 Z\"urich, Switzerland } \email{karlin@lav.mavt.ethz.ch}
\altaffiliation[Also at ]{Institute of Computational Modeling RAS,
Krasnoyarsk 660036, Russia}
\author{ Hans Christian \"Ottinger }
\affiliation {ETH-Z\"urich, Department of Materials, Institute of
Polymers, CH-8093 Z\"urich, Switzerland} \email{hco@mat.ethz.ch}

\begin{abstract}
The grand potential for open systems describes thermodynamics of
fluid flows at low Mach numbers. A  new system of reduced
equations for the grand potential and the fluid momentum is
derived  from the compressible Navier-Stokes equations. The
incompressible Navier-Stokes equations are the quasi-stationary
solution to the new system. It is argued that the grand canonical
ensemble is the unifying concept for the derivation of models and
numerical methods for incompressible fluids, illustrated here with
a simulation of a minimal Boltzmann model in a microflow setup.
\end{abstract}
\date{\today}
\maketitle

The classical incompressible
Navier-Stokes  equation  (INS) is mechanical description of  fluid flows at  low Mach numbers
(Mach number, $\Ma = {U_0}/{c_{\rm s}}$,  is the ratio of the characteristic flow velocity  $U_0$ to the
isentropic sound speed $c_{\rm s}$ defined at some reference temperature $\Tm$ and density $\rhom$).
The INS equation can be  written in  the  Eulerian coordinate system as,
\begin{align}
\label{NS}
\begin{split}
\partial_t u_{\alpha} + u_{\beta} \partial_{\beta}  u_{\alpha}
+ \partial_{\alpha} P &= \frac{1}{\RE}  \partial_{\beta} \partial_{\beta}
u_{\alpha}, \quad
\partial_{\alpha} u_{\alpha} = 0,
\end{split}
\end{align}
where $\bu$ is the  fluid velocity, $P$ is  the
pressure and $\RE$  the
Reynolds number, which characterizes the relative strength of the
viscous and the inertial forces \cite{LandauFM}.
The  pressure  in (\ref{NS}) is
not an independent thermodynamic variable but is rather determined by the condition of the
incompressibility,
\begin{equation}
\label{PIntEq}
\partial_{\beta}\partial_{\beta} P =- \left(\partial_{\beta} u_{\alpha} \right)
\left(\partial_{\alpha} u_{\beta} \right).
\end{equation}
Thus, in order to obtain the pressure at a point, one has to solve
the Laplace equation (\ref{PIntEq}) in a domain, and the
relationship  between the pressure and the velocity becomes highly
nonlocal. The physical meaning of (\ref{PIntEq})  is that in a
system with infinitely fast sound propagation, any pressure (and
thus density) disturbance induced  by the flow is  instantaneously
propagated into the whole domain.

A textbook  justification for the thermodynamics of the INS
description is usually based on the isentropic flow assumption. If
this assumption is valid  for all times (the entropy density is
simply convected by the flow), then the thermodynamic pressure
depends only on the acoustic (density) variations. For low Mach
number flows, these variations adjust to the flow on every spatial
scale in the long time dynamics \cite{Batchlor,LandauFM}. The way
this adjustment takes place is nontrivial, and it was given
considerable attention recently \cite{Majda,NSD,NSBCD}. In
particular,  it was  proved  that weak solutions of the isentropic
compressible Navier-Stokes equations converge to that of the INS
equation for some special boundary conditions (adiabatic absorbing
walls) \cite{NSBCD}.  However, it remains a challenge to give a
thermodynamic derivation of the incompressibility without the
isentropic flow assumption. Such a thermodynamic derivation goes
far beyond academic interest. Indeed, as is well known, the INS
equations are extremely hard to study, both analytically and
numerically. Therefore, an extended system where the flow is
coupled to a dynamic equation for a scalar thermodynamic variable
provides a better starting point for numerical and theoretical
studies of the incompressible hydrodynamics. Indeed, in the
computational fluid dynamics, at least two undeniably successful
routes to avoid the ``elliptic solver problem'', that is, avoiding
the nonlocality of the  pressure
 \eqref{PIntEq}  are well known.
The first is the  so-called artificial compressibility method
introduced by Chorin \cite{Chorin} and Temam \cite{Temam69}, where
an  evolution equation  for the pressure  is postulated instead of
the constraint
 \eqref{PIntEq}  (see e.\ g. \cite{Huges} for a
recent review). The second route is  kinetic-theory  models such
as the lattice Boltzmann method \cite{RMP}, where  a
Boltzmann-like  equation is obtained for  the compressible fluid
flow in the low Mach number limit. The thermodynamics of the
lattice Boltzmann method was clarified \cite{DHT}, and  the method
enjoys a thermodynamically  sound derivation from the Boltzmann
equation \cite{AK5}. Can the compressibility methods be modified
in a way to make them physical models? Furthermore, for emerging
fields of fluid dynamics such as flows at a micrometer scale,
corrections to the incompressibility assumption become crucial
\cite{Karniadakis}.

In this work, we present the thermodynamic description of
incompressible fluid flows. In step one,  we will argue that the
grand potential is the proper thermodynamic potential  to study
the onset of incompressibility. Use of  the grand potential
instead of the entropy
 enormously simplifies the equations of compressible hydrodynamics in the
low Mach number limit. We show that after a short time dynamics
during which acoustic waves are damped  by viscosity, the fast
dynamics of the grand potential becomes singularly coupled to the
slow dynamics of momentum, and  {\it reduced} compressible
Navier-Stokes equations are derived (RCNS). The incompressible
Navier-Stokes equation is the quasi-stationary solution of the
RCNS, when Mach number tends  to zero. RCNS equations are  a
thermodynamically consistent generalization of the compressibility
schemes. Finally, by writing the grand potential for the Boltzmann
equation, we show that, upon an appropriate discretization of
velocities, the present constructions leads to the entropic
lattice Boltzmann method. Correctness of the present
almost-incompressible description is illustrated with a simulation
of a microflow setup.

A low Mach number flow is a setup where only small spatial deviations
of the entropy and the density from the equilibrium value exist.
The grand potential is the natural thermodynamic variable to describe such a setup
(and the corresponding to it  grand canonical ensemble is the natural statistical thermodynamics  framework).
This happens because the balance laws (compressible Navier-Stokes equations or the Boltzmann equation)
are always written for  a sufficiently  small volume element  in the Eulerian frame of reference
(i.e. the volume element is fixed in space). From a thermodynamic standpoint, this volume
element is an open system.
From elementary thermodynamics we know that in an open system,
the thermodynamic equilibrium is conveniently
described in terms of the grand potential $\Omega(\psi, T)$,
where $\psi$ is chemical potential, and $T$ is the temperature.

Now, we shall find the expression for the grand potential in the Eulerian coordinate system for
 a volume element $\delta V$, in  thermodynamic equilibrium.
In the co-moving system, the grand potential is written as
$\Omega_{\rm L}(\psi, T)=-P(\psi, T)\delta V$, where $P$ is the
pressure. The transition to the Eulerian (fixed) system is done by
fixing the momentum, $\Omega_{\rm E}(\psi,T,\bm)=-P\delta
V+\lambda_{\alpha}m_{\alpha}\delta V$, where $\lambda_{\alpha}$
are Lagrange multipliers, and $\bm$ is the momentum density.  For
small values of momentum, the Lagrange multipliers can be
specified by noting that the energy in the Eulerian coordinate
system is $ [\epsilon+(m^2/2\rho)]\delta V$, where $\epsilon$ is
the internal energy density. Using the relationship between the
energy and the grand potential for the thermodynamic equilibrium,
we find that $\lambda_{\alpha}=m_{\alpha}/2\rho+O(m^3)$. Thus, in
the Eulerian coordinate system, the grand potential, up to the
higher-order terms in momentum, is written as,

\begin{equation}
\label{GP} \Omega_{\rm E}=\left(-P+\frac{m^2}{2\rho}\right)\delta
V.
\end{equation}
The difference of the pressure and the kinetic energy is the
(negative of) density of the grand potential, and it will be used
below as the natural thermodynamic potential for the low Mach
number flows:
\begin{equation}
\label{VAR}
{\cal G}= P - \frac{m^2}{2 \rho}.
\end{equation}

Dynamic equations for the set of variables $\rho$, $\bm$ and
${\cal G}$ are written using  the standard compressible
Navier-Stokes equations (CNS) for Newtonian fluids
\cite{LandauFM,Batchlor}. Note that CNS are usually written in
terms of a different set of variables (for example, in terms of
the entropy density $S$ instead of ${\cal G}$). The recomputation
from either form of the CNS equations to the present set of
variables poses no difficulties, and we here write  the final
result:
\begin{widetext}
\begin{align}
\label{NSF}
\begin{split}
\partial_t \,  {\rho} +  \partial_{\alpha}   m_{\alpha}   &=0,\\
\partial_t \;{m_{\alpha}} + \partial_{\beta} \left[\left({\cal G}+ \frac{m^2}{2 \rho}\right) \delta_{\alpha \beta} +
 \frac{m_{\alpha}m_{\beta} }{\rho} +\Pi_{\alpha \beta}
 \right]&=0,\\
\partial_t {\cal G} +\rho\frac{\partial P}{\partial \rho}\biggr|_{S}
\partial_{\alpha}\left(\frac{m_{\alpha}}{\rho} \right)-\partial_{\alpha}
\left[ m_{\alpha}\left(\frac{m^2 }{2 \rho^2} \right)+
\frac{m_{\beta}}{\rho} \Pi_{\alpha \beta}
 \right]&\\+
  \left(1+ \frac{1}{\rho\, \Cv} \kT
\right)\Pi_{\alpha \beta}
 \partial_{\alpha}\left(\frac{ m_{\beta}}{\rho} \right)&=
   \frac{1}{\rho\, \Cv} \kT
 \partial_{\alpha}\left(\kappa \partial_{\alpha} T \right),
\end{split}
\end{align}
\end{widetext}
where $\Pi_{\alpha\beta}$ is the stress tensor of a Newtonian fluid,
 \begin{equation*}
{\Pi}_{\alpha \, \beta } = - \mu   \left[
\partial_{\alpha} \left(\frac{ m_{\beta}}{\rho}\right)  +
    \partial_{\beta}\left(\frac{ m_{\alpha}}{\rho}\right)  -
\delta_{\alpha \beta}\, \left(\frac{2}{D} -\lambda\right)\,
\partial_{\gamma} \left(\frac{m_{\gamma}}{\rho} \right)\right],
\end{equation*}
with  $D$  the spatial dimension, $\mu$  the  shear viscosity, and
$\lambda$ the ratio of bulk to  shear viscosity.  In (\ref{NSF}),
$\kappa$ is the thermal conductivity, $\Cv$ is the specific heat
at constant volume, and the  temperature $T$ is known from the
equation of state. Variations of material parameters such as
viscosity $\mu$ will be ignored in the further discussion. Note
that equations (\ref{NSF}) are ``exact'' in the sense that they
are just the standard CNS equations written for $\rho$, $\bm$, and
the function ${\cal G}$ (\ref{VAR}). However, the physical meaning
of the function  ${\cal G}$ (\ref{VAR}) as the density of the
grand potential is valid only up to the lowest order in momentum,
and thus the form of the CNS  (\ref{NSF}) is an important
intermediate step in  the study of the low Mach number
hydrodynamics.

Since the speed of sound, $c_{\rm s}=\sqrt{\partial P/\partial \rho |_{S}}$, contributes to the
dynamic equation for the grand  potential, it is instructive to rewrite  (\ref{NSF}) as:
\begin{equation}
\label{derivative}
\partial_t {\cal G}+\partial_{\alpha}(c_{\rm s}^2 m_{\alpha})-
\frac{m_{\alpha}}{\rho}\partial_{\alpha}(\rho c_{\rm s}^2)-
\dots
\end{equation}
We expect the following scenario of the onset of incompressibility, as
it can be inferred from (\ref{derivative}):  If the speed of sound  is ``large'', then,
after a ``short-time'' dynamics of the density leading to
$\rho\approx const$,  the third term in (\ref{derivative}) can be
neglected, whereas the second term becomes the dominant contribution to the time derivative of
$ {\cal G}$. The dynamic  equation for the grand potential becomes then singularly perturbed,
and represents the ``fast'' mode coupled to the ``slow'' dynamics of momentum.
The contributions to the time derivative of the grand potential not displayed in (\ref{derivative})
are responsible for corrections to the incompressibility.
We now proceed with quantifying
these statements.

For small deviations from the no-flow situation,
($| { m_{\alpha} m_{\beta}}| \ll P \,{\rho}$, which implies $\Ma \ll 1$),
we are interested in the long time solutions of the CNS equations
(times of the order of the  momentum diffusivity time  $t_{\rm md}\sim \rhom L^2/\mu$,
where $L$ is a characteristic length associated with the flow).
We define a dimensionless number, $\Kn= \mu/(\rho c_{\rm s} L)$,
the ratio of the sound propagation time $L/c_{\rm s}$ and the momentum diffusivity time,  as the
Knudsen number for a general fluid, and  we are considering  $\Kn \ll \Ma\ll 1$.
For the sake of simplicity, we  assume that the Prandtl number  $\Pr \sim 1$ in the subsequent  analysis.
The {\it short time} dynamics
is  isentropic and  linear, and  it is well known that, away from boundaries,  any density perturbation at a distance
$r$ away from the disturbance source decays  as (see \cite{LandauFM}, p.\ 300):

\begin{equation}
\label{damping}
\delta \rho\left(r, t\right) \propto (\La\,  r  L)^{-1/2} \exp{\left( -
\frac{( r -  c_{\rm s}t)^2}{2  \La\,  r  L}\right)}.
\end{equation}
Here  a new dimensionless number $\La$  is defined as:
\begin{equation}
\La = \Kn  \left(2 -\frac{2}{D} +\lambda\right) + \frac{ \Kn
 \left(\gamma -1
  \right)}{\Pr },
\end{equation}
where $\gamma$  the ratio of specific heat at constant pressure
and volume. $\La$ generalizes the notion of Knudsen number for an
arbitrary fluid  (we call it the Landau number  in the honor of
Landau, who explained
    the relevance of this number  in the context
    of acoustic damping \cite{LandauFM}, p. 300).
Thus, the short term hydrodynamics  reveals the following  length
scale $\AKS$ and the  time scale  $t_{\rm a}$ (since we assume
$\Pr\sim 1$, we need not  distinguish between $\La$ and $\Kn$ for
the present purpose):
\begin{equation}
\label{SCALES}
t_{\rm a} \sim \sqrt{\Kn}\left( \frac{L}{c_{\rm s}}\right),\  \AKS \sim  \sqrt{\Kn}L,
\end{equation}
At the time scale larger than $t_{\rm a}$, and on the spatial
scale larger than $\AKS$, the density of the fluid  can be safely
treated as a constant (in the usual isentropic theory,  the
characteristic time for the onset of incompressibility is of the
order $L/c_{\rm s}\gg t_{\rm a}$). Note that the  length scale
$\AKS$  was  also found  in  the derivation of the sub-grid model
from kinetic theory \cite{AKS}. On the time-space scale larger
than (\ref{SCALES}), we can neglect the density variation,  and
the temperature variation  $\delta T$ (from the globally uniform
value $\Tm$) becomes  a function of the grand canonical potential,
\[
 \delta T\approx \frac{\partial T}{\partial P}\biggr|_{\rho}\left( {\cal G}
+\frac{m^2}{2\rho}\right).
\]

Once the time and space scales (\ref{SCALES}) are identified, we
complete the reduction of the CNS equations (\ref{NSF}) by merely
rescaling the variables. The momentum is scaled by the
characteristic momentum $\rhom  U_0$ (known from the initial or
boundary condition),  $\bj=\bm/(\rhom U_0)$, and we introduce the
reduced grand canonical density, $\Theta =  {\cal G} /(\rhom
U_0^2)$. Making  time dimensionless with $t_{\rm a}$, space  with
$\AKS$, ($t\to t/t_{\rm a}$, ${\mathbf{x}}\to {\mathbf{x}}/\AKS$),
neglecting variations of the density, and taking into account the
thermodynamic relation for the temperature mentioned above, the
two last equations in  \eqref{NSF}  reduce to the following
scale-independent closed set of  equations for the dimensionless
grand potential  and momentum:
\begin{widetext}
\begin{eqnarray}
\label{MINS}
\begin{aligned}
\partial_t j_{\alpha}  &=- \Ma \, \partial_{\beta} \Biggl[ j_{\alpha} j_{\beta}
     + \delta_{\alpha \beta }\left(\Theta + \frac{j^2}{2}
\right)\Biggr]
 +\sqrt{\Kn}
\left(
 1+ \lambda  -
 \frac{2}{D}\right)\partial_{\alpha}\partial_{\beta}
{j_{\beta} }
 +
\sqrt{\Kn}
\partial_{\beta}\partial_{\beta}
{ j_{\alpha}}, \,
\\
\partial_t \Theta
&= -\frac{1}{\Ma }\partial_{\alpha} \left[ j_{\alpha}\left( 1-
\frac{\,\Ma^2j^2  }{2 } \right)
 \right]+\sqrt{\Kn}
 \Biggl[\frac{\gamma }{\Pr}
 \partial_{\alpha}\partial_{\alpha}
\Theta
+ \left(\frac{\gamma }{\Pr} - 1 \right)
 \partial_{\alpha}\partial_{\alpha}
\frac{j^2}{2}
+\left(\partial_{\alpha}j_{\beta}
\right)\left(\partial_{\alpha}j_{\beta} \right)
 \\
& +  \kTmean \frac{1}{2\, \Cv} \left( \partial_{\alpha} j_{\beta}
+
    \partial_{\beta}  j_{\alpha}\right)\left( \partial_{\alpha} j_{\beta} +
    \partial_{\beta}  j_{\alpha}\right)
-
\left(
 1+ \lambda  -
 \frac{2}{D}\right)
 \partial_{\beta}\left( j_{\beta} \partial_{\alpha}  j_{\alpha}\right)
\Biggr].
\end{aligned}
\end{eqnarray}
\end{widetext}
Note that all ``material parameters'' appearing in (\ref{MINS}) ($\Cv$, $\gamma$,  $\lambda$, $\kappa$)
are evaluated
at equilibrium at $\rhom$ and $\Tm$.

The  {\it reduced} set of compressible Navier-Stokes equations
(\ref{MINS})  is valid for   $\Kn \ll \Ma \ll 1$, and, as we
explained above, on the scales larger than acoustic scales
(\ref{SCALES}). Roughly speaking, (\ref{MINS}) is what survives
from the compressible Navier-Stokes equations  just before the
incompressibility sets on. Indeed, the time derivative of $\Theta$
becomes singularly perturbed as the Mach number tends to zero, and
we  recover the incompressibility condition, $\partial_{\alpha}
j_{\alpha} = 0$, as the quasi-stationary solution of the system
(\ref{MINS}). This solution, when substituted in the momentum
equation, recovers the INS equation (\ref{NS}) with the usual
accuracy  of the order $O(\Ma^2)$. Note that the velocity in the
INS equations recovered from (\ref{MINS}) is $\bu=\bj/\rhom$.
 Corrections to the quasi-stationary solution
can be found in a systematic way (see, e.\ g.\
\cite{PhRep,GKbook}), and we do not address this here. The
following point needs to be stressed: The dissipation terms
(proportional to $\sqrt{\Kn}$) {\it cannot} be neglected  in the
equation for the grand potential (\ref{MINS}) and simultaneously
kept in the momentum equation. This is at variance with the
artificial compressibility method \cite{Chorin,Temam69}.  In other
words, the RCNS is {\it the minimal thermodynamic system} for
incompressible hydrodynamics.

In this Letter, we reported a new basic physical fact of fluid
dynamics: Grand potential (\ref{GP}) for low Mach number flows
gives the thermodynamic description of the incompressible
phenomena. The resulting system (\ref{MINS}) includes a {\it
local} (non-advected) equation for the scalar thermodynamic field.
What follows from this fact? Let us list some of the consequences:

\begin{itemize}

\item With the corrections mentioned above, the numerical schemes
of the artificial compressibility kind become a firm status of
physical models.

\item The structure of the coupling between the flow and the
thermodynamic variable hints at that the true incompressible flows
are attractors of system (\ref{MINS}), with the INS as the
leading-order approximation. Because of the singular perturbation
nature, it may be easier to study attractors of (\ref{MINS}),
rather than of the INS equations.

\item System (\ref{MINS}) can be a starting point for a systematic
derivation of nonlinear models for heat transport in the
nearly-incompressible fluids such as multi-phase fluids, polymeric
liquids and melts etc (this is relevant even in the linear case,
see, e.\ g.\ \cite{Bechtel03}).

\item In the celebrated Kolmogorov theory, equation for the
kinetic energy is used to make predictions about the structure of
the fully developed turbulence. The present thermodynamic approach
 unambiguously delivers the density of the grand potential as the scalar
field associated with the incompressible fluid flow, and thus can
be relevant to develop theories of turbulence through studying the
resulting balance equation.

\end{itemize}


Let us dwell on the use of (\ref{MINS}) for numerical simulation
of incompressible flows. Recall that the spectral methods
\cite{Orszag77} for the INS (\ref{NS}) based on the Fast Fourier
Transform (FFT) for solving the Laplace equation (\ref{PIntEq})
are very efficient for high Reynolds number flows in simple
geometries.
 On the other hand, thanks to a relatively simple structure of
the system (\ref{MINS}) (the equation for the scalar variable
contains no convected derivative), it can be addressed by a host
of discretization methods (see e.\ g.\ \cite{Huges}). The system
(\ref{MINS}) can be useful for simulation of flows in complex
geometries and especially non-stationary problems, that is, in the
situations where one seeks to avoid solving the Laplace equation
by restoring to some relaxation schemes (artificial
compressibility, lattice Boltzmann etc).
Special attention should be payed to the fact that the system
(\ref{MINS}) contains terms of different order of magnitude. This
situation is typical for all relaxational schemes such as, for
example, the lattice Boltzmann method, and we do not discuss here
how to deal with this issue numerically (see, e.\ g.\
\cite{RMP,DHT,AK5}). Note, however, the important smoothing effect
of the diffusion term (inversely proportional to the Prandtl
number) in the equation for $\Theta$ which makes the physical
system (\ref{MINS}) more amenable to numerics, that is, less
numerically stiff than artificial compressibility methods. These
numerical issues remain out of the scope of the present Letter,
and will be addressed in a separate publication.

\begin{figure}
\includegraphics[scale=0.45]{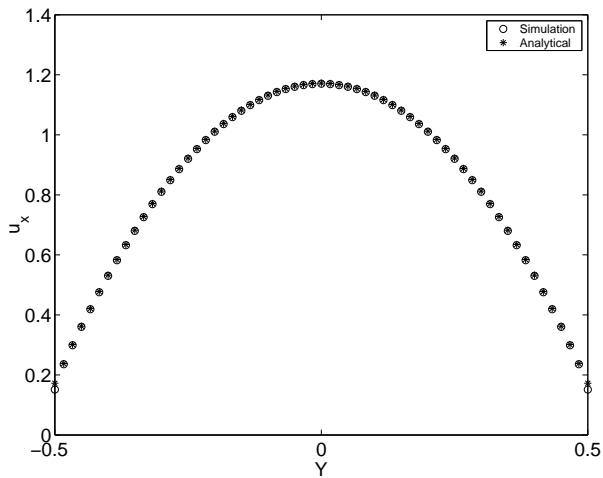}
\caption{\label{Fig1} Velocity  profile of the $2D$ body force driven
Poiseuille flow  at $\Kn=0.035$ and $\Ma=0.01$.}
\end{figure}

While the new system of reduced compressible Navier-Stokes
equations (\ref{MINS}) is our central result, we conclude this
Letter with a more general statement that the grand potential (and
the relevant grand canonical ensemble)
 can be implemented
for eventually  any more microscopic setup (and  not only for
compressible Navier-Stokes equations as above). This viewpoint
provides a unified setting for derivations of a variety of
mesoscopic or molecular dynamics models for numerical simulation
of incompressible and nearly-incompressible flows. As an
illustration here, let us assume Boltzmann's description with the
one-particle distribution function $f(\bv,\bx,t)$. Starting from
the general form of grand potential, ${\cal
G}(f,\alpha,\beta,\blambda)= \int f\left[
\ln{f}+\alpha+\lambda_{\alpha}v_{\alpha}+\beta v^2\right]d\bv$, it
is easy to show that at equilibrium $f^{\rm
eq}(\bv,\alpha,\beta,\blambda)$ (defined from $\delta {\cal
G}=0$), we have ${\cal  G}(f^{\rm eq})={\cal  G}^{\rm
eq}(\alpha,\beta)+(\lambda^2/2) \int f^{\rm
eq}(\bv,\alpha,\beta,0)(v^2/2)d\bv$ for small $\lambda$. When the
velocity integral  ${\cal G}(f,\alpha,\beta,\blambda)$ is
evaluated  with the Gauss-Hermit quadrature with the weight
$\exp(\beta v^2)$ at fixed $\beta$, one obtains the entropy
function of the lattice Boltzmann method  \cite{DHT,AK5}.
 The
present alternative derivation based on the grand potential is
new. 
Thus, the entropic lattice Boltzmann method is a valid physical
model for nearly-incompressible flows, and can be used for finite
but small Knudsen low Mach number flow problems often encountered
in the microflows \cite{Karniadakis}. In Fig. \ref{Fig1} we
present excellent comparison between  the analytical solution to
the Bhatnagar-Gross-Krook (BGK) kinetic equation \cite{Cerci} and
the entropic lattice Boltzmann simulation for the body-force
driven $2D$ Poiseuille microflow.
 A detailed study of microflows within the entropic lattice
Boltzmann setting is presented elsewhere \cite{Ansumali04}.

While this paper was in the revision process, we learned about a
very recent paper \cite{Hallatschek04} which states the usefulness
of the grand potential in the context of plasma turbulence
simulations. Although \cite{Hallatschek04} does not deal with the
incompressibility per se, it indicates that grand potential may be
a relevant thermodynamic variable in flowing systems way beyond
``ordinary'' fluids.

{\it Acknowledgments.} Discussions with Alexander Gorban were
important starting point of this work. Useful comments of C.\
Frouzakis, M.\ Grmela, V.\ Kumaran, and A.\ Tomboulides are kindly
acknowledged. IK and SA were supported by the Swiss Federal
Department of Energy (BFE) under the project Nr. 100862 ``Lattice
Boltzmann simulations for chemically reactive systems in a
micrometer domain".

\bibliography{inc}

\end{document}